\documentclass[sigconf,10pt]{acmart}

\usepackage{algorithm}
\usepackage{algorithmic}
\usepackage{subcaption}
\usepackage{xspace}
\usepackage{makecell}

\AtBeginDocument{%
  }

\begin{document}

\setcopyright{cc}
\acmYear{2026}\copyrightyear{2026}
\acmConference[APNet 2026]{The 10th Asia-Pacific Workshop on Networking}{August 06--07, 2026}{Singapore, Singapore}
\acmBooktitle{The 10th Asia-Pacific Workshop on Networking (APNet 2026), August 06--07, 2026, Singapore, Singapore}
\acmDOI{10.1145/3820441.3820455}
\acmISBN{979-8-4007-2664-4/26/08}

\title{EdgePoW: Adaptive Ingress-Aware Defense with Non-Interactive PoW Against Volumetric SYN Floods}

\author{Wenyang Jia}
\email{wjia45@gatech.edu}
\affiliation{%
  \institution{ICN Lab, Shenzhen Graduate School, Peking University}
  \city{Shenzhen}
  \country{China}}

\author{JingJing Wang}
\email{jingjing\_wang@pku.edu.cn}
\affiliation{%
  \institution{ICN Lab, Shenzhen Graduate School, Peking University}
  \city{Shenzhen}
  \country{China}}

\author{Xianneng Zou}
\email{sagezou@tencent.com}
\affiliation{%
  \institution{Tencent}
  \city{Shenzhen}
  \country{China}}

\author{Kai Lei}
\authornote{Corresponding author. Email: leilk@pkusz.edu.cn}
\email{leilk@pkusz.edu.cn}
\affiliation{%
  \institution{ICN Lab, Shenzhen Graduate School, Peking University}
  \city{Shenzhen}
  \country{China}}

\begin{abstract}
The stability of Internet services is persistently challenged
by large volumetric TCP SYN floods, for which conventional
defenses such as SYN Cookies preserve server state but still
amplify bandwidth pressure. This paper presents EdgePoW,
an ingress aware defense architecture that integrates
non interactive Proof of Work with an SDN control plane for
managed edge networks. The controller monitors per ingress
SYN pressure and raises PoW difficulty when flooding is
detected. If traffic mainly originates from a stable source
region, enforcement is refined to the offending source prefix
to reduce overhead on benign co located clients; otherwise,
ingress wide enforcement is retained under randomized or
spoofed sources. We further design a conservative Difficulty
Discovery Protocol that reuses TCP retransmissions and
commits difficulty updates only after a successful handshake.
Experiments on a custom SDN testbed show restored application
QoS under concentrated and spoofed floods, 11.7\%
higher benign client throughput than ingress only enforcement,
and below 0.8\% transient false escalations under 2\%
random loss.
\end{abstract}

\begin{CCSXML}
<ccs2012>
<concept>
<concept_id>10003033.10003068.10003073</concept_id>
<concept_desc>Networks~Denial-of-service attacks</concept_desc>
<concept_significance>500</concept_significance>
</concept>
<concept>
<concept_id>10003033.10003068</concept_id>
<concept_desc>Networks~Network security</concept_desc>
<concept_significance>300</concept_significance>
</concept>
<concept>
<concept_id>10003033.10003083</concept_id>
<concept_desc>Networks~Programmable networks</concept_desc>
<concept_significance>300</concept_significance>
</concept>
</ccs2012>
\end{CCSXML}

\ccsdesc[500]{Networks~Denial-of-service attacks}
\ccsdesc[300]{Networks~Network security}
\ccsdesc[300]{Networks~Programmable networks}

\keywords{Network Security, Denial-of-Service (DoS), Software-Defined Networking (SDN)}

\maketitle
\vspace*{-1.05\baselineskip}

%%
%% 1. INTRODUCTION
%%
\section{Introduction}

The Internet's open design leaves it vulnerable to volumetric
DoS attacks. With attack rates now reaching Tbps~\cite{cloudflare2024,yoachimik2024q1,yoachimik2024q4,jia2025blocksdnvc},
traditional defenses are increasingly inadequate. SYN
Cookies~\cite{pan2024tcp,eddy2007syn} mitigate server state exhaustion but still trigger
a SYN-ACK per SYN, amplifying bandwidth depletion under
massive floods.

This paper introduces EdgePoW, a framework that
integrates miniature Proof-of-Work (PoW) challenges into
TCP SYN packets, governed by an adaptive in-network control
loop. Our approach meets three critical criteria for an ideal
defense:
\begin{enumerate}
  \item it shifts the computational burden from the victim to
    the attacker;
  \item it enables proactive filtering deep within the network
    fabric, neutralizing threats before they reach their
    target; and
  \item it provides adaptive response, applying stringent
    validation only when and where needed.
\end{enumerate}

The key design idea in EdgePoW is an adaptive ingress-aware
control loop. The controller first detects stressed ingress
edges in real time and raises PoW difficulty there; it then
refines the policy to a source prefix only when the ingress
exhibits a stable dominant prefix. This ingress-first,
prefix-refined control loop keeps the system effective under
spoofed floods while preserving selectivity when source
identities are trustworthy. A key open problem
in any variable-threshold PoW system is how clients
discover the required difficulty~\cite{delaughter2025syn}; we address this with a
Difficulty Discovery Protocol (DDP) that leverages TCP's
retransmission mechanism for transparent adaptation.

The principal contributions of this work are:
\begin{itemize}
  \item \textbf{Adaptive ingress-aware PoW control.} We
    address the static-threshold limitation of SYN PoW~\cite{delaughter2025syn}
    by introducing a controller that continuously senses
    per-ingress SYN anomalies, installs ingress-wide PoW
    rules for spoofed floods, and refines them to
    $(\mathit{ingress},\mathit{prefix})$ granularity when source
    concentration is stable.
  \item \textbf{A robust Difficulty Discovery Protocol (DDP).}
    We solve the open problem of how clients learn the
    current PoW difficulty~\cite{delaughter2025syn} with a lightweight protocol
    that piggybacks on TCP retransmission, performs
    connection-local tentative escalation on timeouts,
    and commits values only after a successful handshake.
    Transient false escalations stay below 0.8\% under 2\%
    random loss.
  \item \textbf{Empirical validation on a physical SDN testbed.}
    EdgePoW restores application-level QoS during
    spoofed SYN floods, and prefix refinement improves
    benign-client throughput by 11.7\% over ingress-only
    enforcement while preserving attack suppression.
\end{itemize}

%%
%% 2. BACKGROUND
%%
\section{Background}

TCP SYN floods exploit the three-way handshake: attackers
send massive volumes of SYN packets, exhausting bandwidth
and server state~\cite{eddy2007syn,eddy2022tcp}. SYN Cookies~\cite{pan2024tcp} mitigate
state exhaustion by encoding connection metadata in the
SYN-ACK, but since every SYN still elicits a response, they
amplify egress load under Tbps-scale floods~\cite{delaughter2025syn,yoo2024smartcookie}.
Proof-of-Work (PoW), originally proposed for spam
mitigation~\cite{dwork1992pricing}, imposes a per-request computational cost
that is negligible for legitimate senders but prohibitive at
high rates. Prior TCP-PoW proposals used interactive
challenge--response models~\cite{juels1999client,noureddine2019revisiting,scholz2020}
that still trigger server transmissions; we adopt a
non-interactive design embedding the PoW in the initial
SYN~\cite{delaughter2025syn}. In managed edge networks, a
programmable control plane provides the visibility needed to
make this defense adaptive: the controller detects stressed
ingress edges and pushes ingress-wide or prefix-refined PoW
rules.

\paragraph{Motivation.}
DeLaughter and Sollins~\cite{delaughter2025syn} demonstrated the first
non-interactive SYN PoW prototype with eBPF-based
verification at line rate. However, their static threshold
($k=16$) applied uniformly to all traffic causes two problems:
(1)~peacetime connections incur unnecessary overhead;
(2)~a global threshold cannot differentiate attacked ingresses
from clean ones. We address both with SDN-driven ingress
detection and opportunistic prefix refinement.

%%
%% 3. DESIGN AND IMPLEMENTATION
%%
\section{EdgePoW Design and Implementation}

The system has two components: client-side PoW generation
in the TCP stack and network-side verification managed by an
SDN controller. The controller reasons over policy
domains---an ingress edge or an $(\mathit{ingress},\mathit{src\_prefix})$
pair---through three phases: (1)~monitoring per-ingress SYN
counters; (2)~detection via Algorithm~\ref{alg:controller}; and
(3)~enforcement of elevated PoW rules. Clients discover new
difficulty via the Difficulty Discovery Protocol (\S\ref{sec:ddp}).

\subsection{Design Overview}

EdgePoW embeds a non-interactive proof-of-work in each
TCP SYN packet, with the difficulty level set dynamically by
the SDN controller based on real-time traffic conditions.
During peacetime, the default difficulty is minimal (or zero),
imposing no perceptible cost on legitimate connections. When
the controller detects a volumetric SYN flood at a specific
ingress, it raises the difficulty for that ingress, forcing
attackers to spend computational effort per SYN while the
rest of the network remains unaffected.

\paragraph{Client-Side PoW Generation.}
A compliant client hashes the SYN header fields
(source/destination addresses, ports, and a nonce) until the
resulting hash meets the configured difficulty; at low
difficulty, this cost is negligible for legitimate connections
but limits high-rate attackers whose per-SYN computation must
scale linearly with their sending rate.

\paragraph{Network-Side Verification.}
Difficulty $d$ defines the required number of leading zero
bits in the hash output; the expected iterations to find a
valid nonce are $k = 2^d$. The controller tunes $d$ per policy
domain---an entire ingress during spoofed floods, or a
specific source prefix when stable. Clients discover the
current $d$ via the Difficulty Discovery Protocol (\S\ref{sec:ddp}).

\subsection{PoW Implementation Details}

\paragraph{Nonce Encoding.}
The nonce must reside in a TCP header field modifiable
without violating semantics. We consider the 32-bit Sequence
Number (whose ISN must be unpredictable) and the 32-bit
Acknowledgment Number (unused in SYNs and typically zero);
we adopt the latter to encode the nonce, simplifying
implementation without interacting with kernel connection
state.

\paragraph{Hash Function Selection.}
Choosing the hash function $H$ balances verification cost and
security: it must be fast enough for line-rate data-plane
checks yet resistant to nonce prediction beyond brute force.
Our prototype uses SuperFastHash~\cite{hsieh} for simplicity;
production deployments should adopt a standardized
cryptographic hash (e.g., truncated SHA-256). The framework
is hash-agnostic provided clients and verifiers use the same
$H$.

\paragraph{Hash Inputs.}
{\sloppy To ensure integrity and per-connection uniqueness, the hash
input binds source/destination IP addresses,
source/destination TCP ports, and the nonce. This prevents
replay across flows or spoofed sources; adding a
coarse-grained timestamp further limits validity to a short
window and mitigates precomputation.\par}

\subsection{The SDN Controller: Dynamic and Targeted Defense}

The controller continuously monitors per-ingress SYN rates
and, when a flood is detected, installs PoW rules scoped to
the narrowest safe policy domain.

\paragraph{Real-time Threat Detection.}
The controller maintains per-ingress and per-prefix SYN
baselines and tracks the dominant-prefix share
\[
  \rho(e) = \max_{S} \frac{\mathit{syn\_rate}[e,S]}{\mathit{syn\_rate}[e]},
\]
for each ingress $e$.

\paragraph{Dynamic Threshold Management.}
The controller manages a global default $d_\mathit{default}$
(e.g., $d=0$) for peacetime and elevated $d_\mathit{attack}$
(e.g., $d=16$) during floods.

\subsubsection{Targeted Policy Enforcement}

Upon detecting a flood at ingress $e$, the controller installs
an ingress-wide rule:
\begin{quote}\small
  Match: $\{\mathit{ingress} = e,\ \mathit{tcp.flags} = \mathrm{SYN}\}$ \\
  Action: $\{\mathit{verify\_pow}(d_\mathit{attack})\}$
\end{quote}
If a stable dominant prefix $S$ is observed (high $\rho(e)$
and persistent excess from $S$), the controller refines to:
\begin{quote}\small
  Match: $\{\mathit{ingress} = e,\ \mathit{ip.src} = S,\ \mathit{tcp.flags} = \mathrm{SYN}\}$ \\
  Action: $\{\mathit{verify\_pow}(d_\mathit{attack})\}$
\end{quote}

Algorithm~\ref{alg:controller} dynamically adjusts difficulty
and chooses the narrowest safe policy domain. We evaluate
sensitivity of $\theta$ and $\tau_\mathit{detect}$ in
\S\ref{sec:sensitivity}.

\begin{algorithm}[t]
\caption{Adaptive PoW Controller Policy}
\label{alg:controller}
\begin{algorithmic}[1]
\STATE \textbf{Inputs:} \textit{syn\_rate}[edge], \textit{syn\_rate}[edge,prefix],
  \textit{baseline}[edge], \textit{baseline}[edge,prefix]
\STATE \textbf{Parameters:} $d_\mathit{default}$,
  $d_\mathit{attack\_range} = [d_\mathit{min}, d_\mathit{max}]$,
  $T_\mathit{budget}$, $\rho_\mathit{refine}$, $\theta_p$,
  $\tau_\mathit{refine}$
\FOR{each ingress $e$ every $\Delta t$}
  \IF{$\mathit{syn\_rate}[e] > \theta \cdot \mathit{baseline}[e]$
        for $\tau_\mathit{detect}$}
    \STATE Find minimal $d^* \in [d_\mathit{min}, d_\mathit{max}]$ such that
    \STATE \quad $\mathit{early\_drop\_expected}(d) \ge 95\%$ and
      $T_\mathit{conn}(d) = 2^d/H \le T_\mathit{budget}$
    \STATE Install coarse rule: $\mathit{edge}(e)$, match (SYN),
      action $\mathit{verify\_pow}(d^*)$
    \STATE $S^* \leftarrow \arg\max_{S}\,\mathit{syn\_rate}[e,S]$
    \IF{$\rho(e) \ge \rho_\mathit{refine}$ and
          $\mathit{syn\_rate}[e,S^*] > \theta_p \cdot \mathit{baseline}[e,S^*]$
          for $\tau_\mathit{refine}$}
      \STATE Replace coarse rule with refined rule:
      \STATE \quad $\mathit{edge}(e)$,
        match ($\mathit{src\_prefix} = S^* \wedge \mathrm{SYN}$),
        action $\mathit{verify\_pow}(d^*)$
    \ENDIF
  \ELSIF{$\mathit{syn\_rate}[e] < \beta \cdot \mathit{baseline}[e]$
           for $\tau_\mathit{clear}$}
    \STATE Retract rules on $e$ \quad\{hysteresis to avoid flapping\}
  \ENDIF
\ENDFOR
\STATE \textbf{Notes:} Ingress-wide enforcement is the safe default;
  refine only when source concentration is stable.
\end{algorithmic}
\end{algorithm}

\subsection{Client Difficulty Discovery Protocol}
\label{sec:ddp}

A key challenge is enabling clients to learn the current PoW
difficulty~\cite{delaughter2025syn}. Prior proposals (DNS advertisement, NACKs,
AIMD probing) each add overhead or complexity. Our DDP
piggybacks on TCP's SYN retransmission: a timeout is an
ambiguous signal (loss or stronger policy), so DDP performs
only tentative escalation and writes a new difficulty to
cache only after a SYN-ACK confirms it.

\subsubsection{Protocol Design}

In Phase~1 (\emph{Tentative Escalation}), a client initiates
at its last confirmed difficulty (or $d_\mathit{default}$).
On timeout, it increments a connection-local trial difficulty
by step $\Delta$ and retries up to $R_\mathit{max}$ times;
this tentative state is discarded if the connection fails.
In Phase~2 (\emph{Confirmed Caching}), the new difficulty is
written to cache only after a successful SYN-ACK. Cache
entries expire after $T_\mathit{cache}$; stale high-difficulty
entries remain safe as they satisfy any lower threshold. See
Algorithm~\ref{alg:ddp}.

\begin{algorithm}[t]
\caption{Client-Side Difficulty Discovery Protocol}
\label{alg:ddp}
\begin{algorithmic}[1]
\STATE \textbf{Parameters:} $d_\mathit{default}$, $\Delta$ (step size),
  $R_\mathit{max}$ (max retries), $T_\mathit{syn}$ (SYN timeout),
  $T_\mathit{cache}$ (cache TTL)
\STATE $d_\mathit{confirmed} \leftarrow
  \mathit{cache}[\mathit{dst\_prefix}]$ if valid, else $d_\mathit{default}$
\STATE $d \leftarrow d_\mathit{confirmed}$
  \quad\{connection-local tentative difficulty\}
\STATE $r \leftarrow 0$
\REPEAT
  \STATE Generate SYN with PoW at difficulty $d$
  \STATE Send SYN; wait up to $T_\mathit{syn}$ for SYN-ACK
  \IF{SYN-ACK received}
    \STATE $\mathit{cache}[\mathit{dst\_prefix}] \leftarrow
      (d,\ \mathit{now} + T_\mathit{cache})$
    \STATE \RETURN Connected
  \ELSE
    \STATE $r \leftarrow r + 1$;\quad
      $d \leftarrow \min(d + \Delta,\ d_\mathit{max})$
  \ENDIF
\UNTIL{$r \ge R_\mathit{max}$}
\STATE Preserve existing cache entry; \RETURN Failure
\end{algorithmic}
\end{algorithm}

\subsubsection{Convergence and Design Rationale}

Retries to discover $d_\mathit{attack}$ is
$N_\mathit{retry} = \lceil (d_\mathit{attack} - d_\mathit{default}) / \Delta \rceil$,
with latency bounded by $N_\mathit{retry} \times T_\mathit{syn}$.
With $\Delta=4$ and $T_\mathit{syn}=1$\,s, a client discovers
$d_\mathit{attack}=16$ in at most four retries (4\,s).
Timeout-driven escalation is tentative: random loss may cause
a higher $d_\mathit{probe}$, but it is never cached unless the
handshake succeeds (\S\ref{sec:ddp-robustness}). Key
properties:
\begin{itemize}
  \item \textbf{Zero extra bandwidth:} failed SYNs trigger no
    server-side response.
  \item \textbf{Self-correcting:} cache expiry de-escalates
    after an attack clears; stale high-$d$ entries are always
    accepted by lower thresholds; failures never overwrite
    confirmed values.
  \item \textbf{TCP-compatible:} DDP requires no new packet
    types or protocol extensions.
\end{itemize}

%%
%% 4. METRICS AND METHODOLOGY
%%
\section{Metrics and Methodology}

Following the dual-control methodology of~\cite{delaughter2023thesis,delaughter2022context},
we derive Efficacy $E$ (QoS recovered) and Overhead $O$
(peacetime cost) from four experiment types per
configuration. QoS is measured as transaction rate:
successful 1\,KB HTTPS GETs/s (1\,s timeout, 120\,s runs).

\begin{figure}[t]
  \centering
  \includegraphics[width=\linewidth]{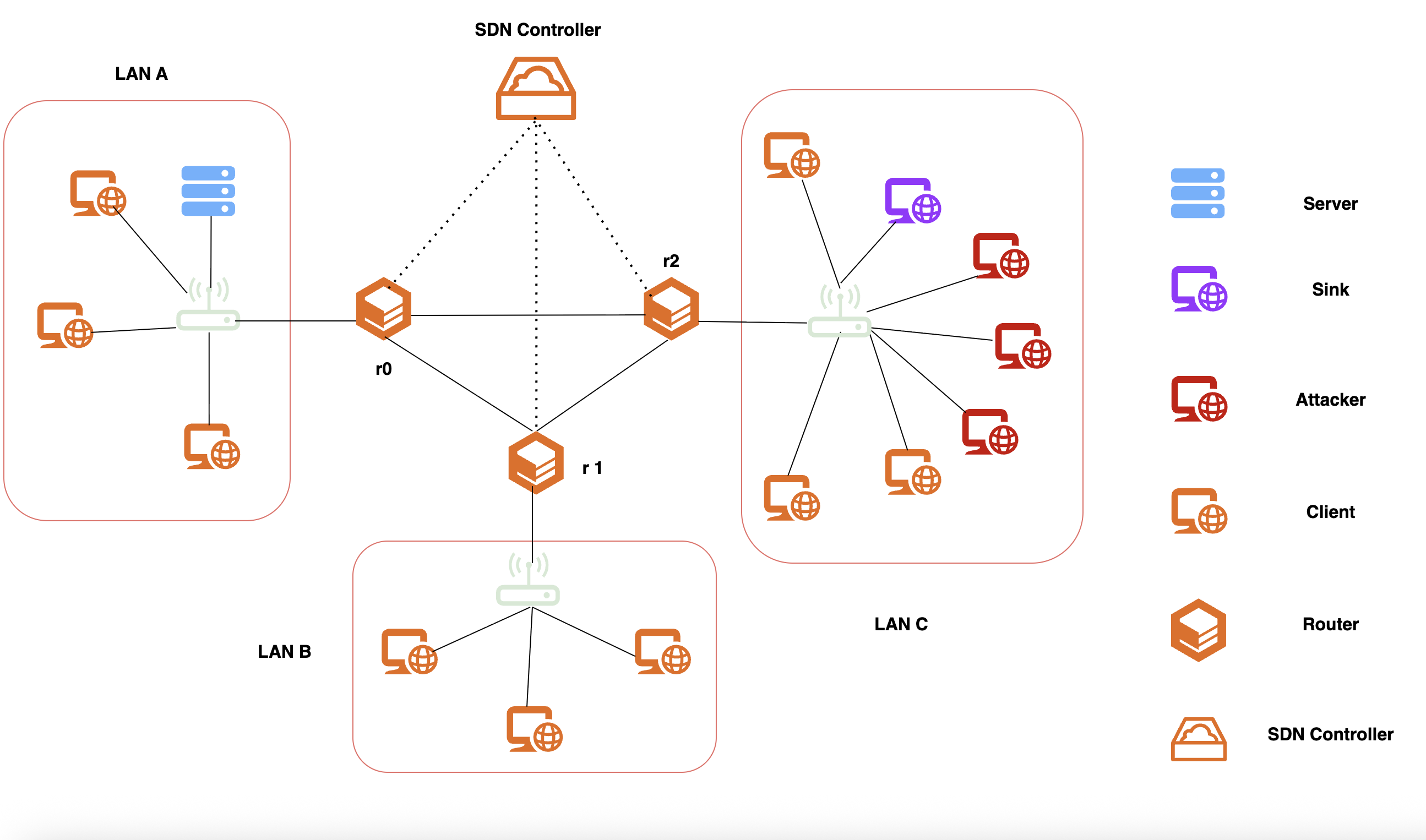}
  \caption{Experimental topology (clients, attackers,
    SDN-enabled routers).}
  \label{fig:topology}
\end{figure}

\paragraph{Testbed.}
Experiments ran on a physical testbed (Fig.~\ref{fig:topology})
with a central server, nine clients, and three attackers across
three LANs. SDN-enabled routers (r0--r2) are coordinated by
a Ryu controller (OpenFlow~1.3) that polls counters every
500\,ms. LAN~A is co-located with the server; LAN~B traverses
the core; LAN~C shares links with attack nodes (most
contested) and is split into attacker and benign /24 prefixes
behind $r_2$. All devices run Ubuntu~22.04 with Open
vSwitch~2.17. The controller triggers ingress-wide protection
after two anomalous windows and refines to a prefix when it
contributes $\ge 85\%$ of the ingress SYN volume for 1\,s.

\paragraph{Attack variants.}
We test three SYN floods representing escalating severity:
(i)~a low-rate curl-based flood, (ii)~a high-rate C-based
generator ($>350$\,Kpps, modelled after Mirai~\cite{mirai})
launched from a stable LAN~C prefix, and (iii)~the same
high-rate generator with fully randomized source IP addresses.
Each attack variant is tested under four defense
configurations: No-Defense, SYN Cookies, static SYN-PoW
($d=16$ globally), and EdgePoW ($d_\mathit{default}=4$,
$d_\mathit{attack}=16$). Under stable-source floods, the
controller refines to the attacker /24 after $0.8 \pm 0.2$\,s;
under spoofing, the ingress-wide rule persists because no
single prefix dominates.

\paragraph{DDP robustness tests.}
We ran 10,000 connection attempts per scenario to evaluate DDP
resilience: random packet loss at 1\% and 2\% rates, a
$4{\times}$ flash crowd to stress the system, stale-low cache
($d=4 \to 16$) simulating attack onset, stale-high cache
($d=16 \to 4$) simulating attack clearance, and
server-unreachable conditions (route withdrawn for 60\,s).
These scenarios test whether DDP converges correctly and
avoids cache corruption.

%%
%% 5. RESULTS AND ANALYSIS
%%
\section{Results and Analysis}

\subsection{Baseline, Overhead, and Unmitigated Threat}

\begin{figure*}[t]
  \centering
  \begin{subfigure}[b]{0.48\textwidth}
    \includegraphics[width=\textwidth]{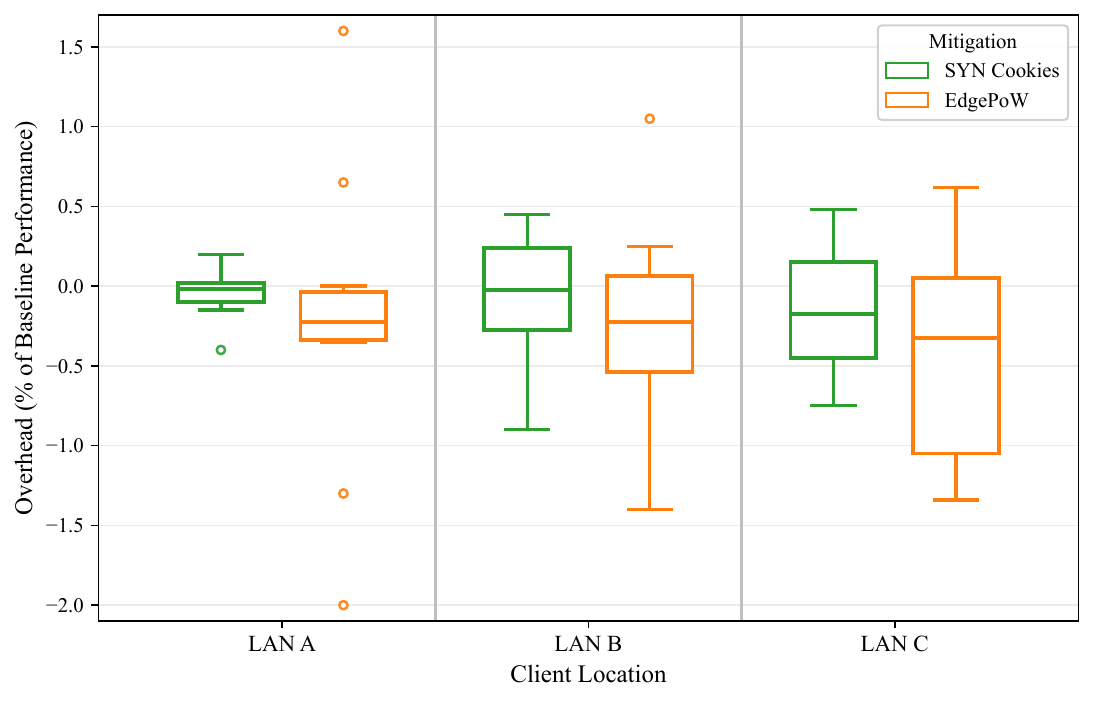}
    \caption{Peacetime overhead ($< 2\%$ for both defenses).}
  \end{subfigure}
  \hfill
  \begin{subfigure}[b]{0.48\textwidth}
    \includegraphics[width=\textwidth]{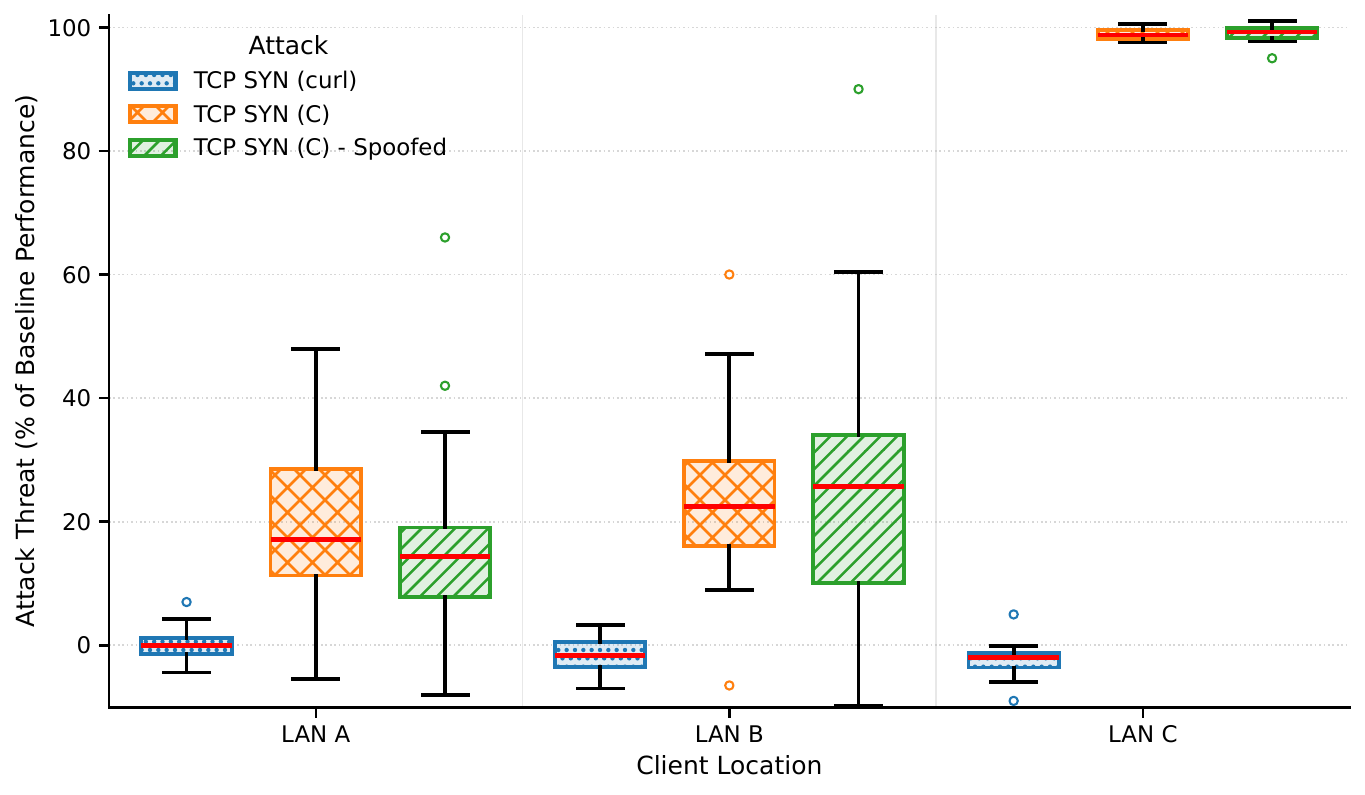}
    \caption{Unmitigated 350\,Kpps flood: LAN~C full DoS.}
  \end{subfigure}
  \\[4pt]
  \begin{subfigure}[b]{0.48\textwidth}
    \includegraphics[width=\textwidth]{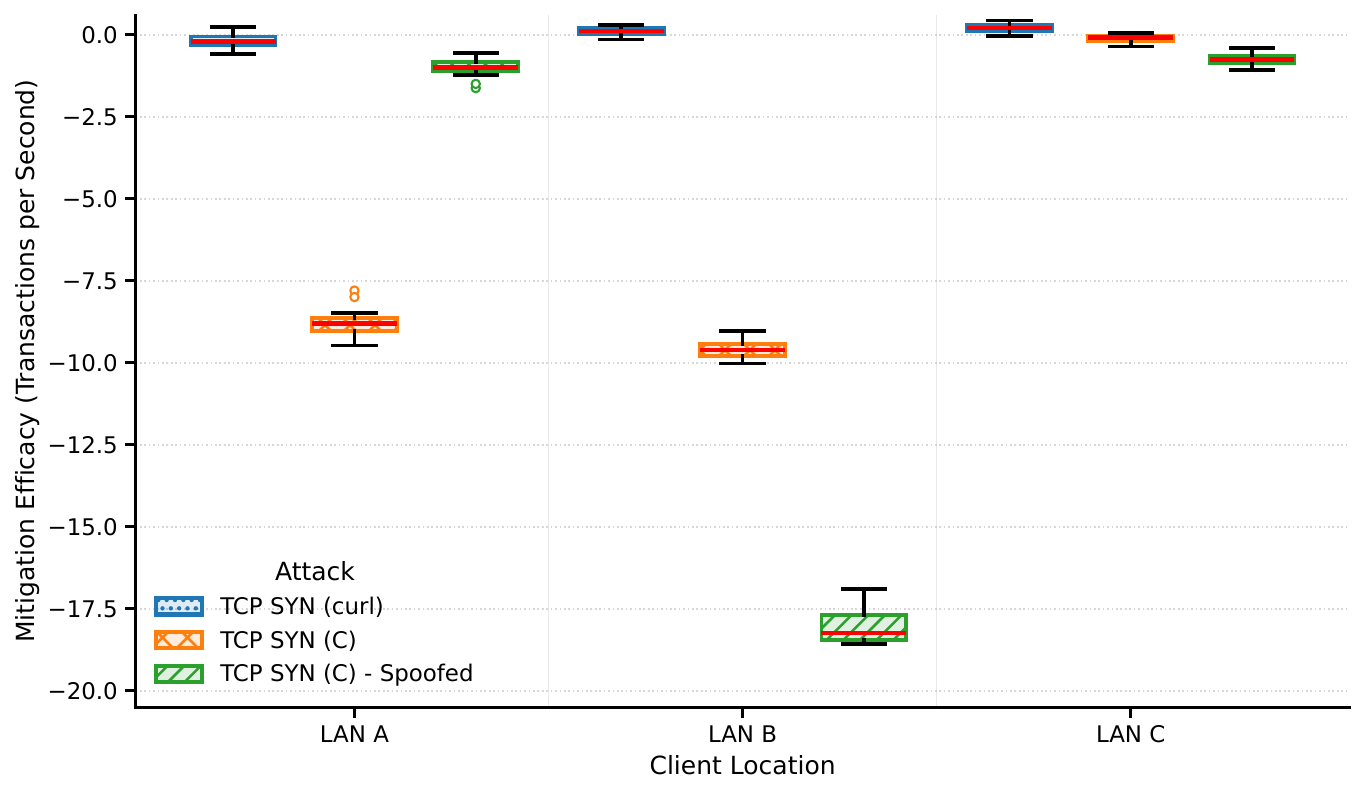}
    \caption{SYN Cookies amplify congestion (negative efficacy).}
  \end{subfigure}
  \hfill
  \begin{subfigure}[b]{0.48\textwidth}
    \includegraphics[width=\textwidth]{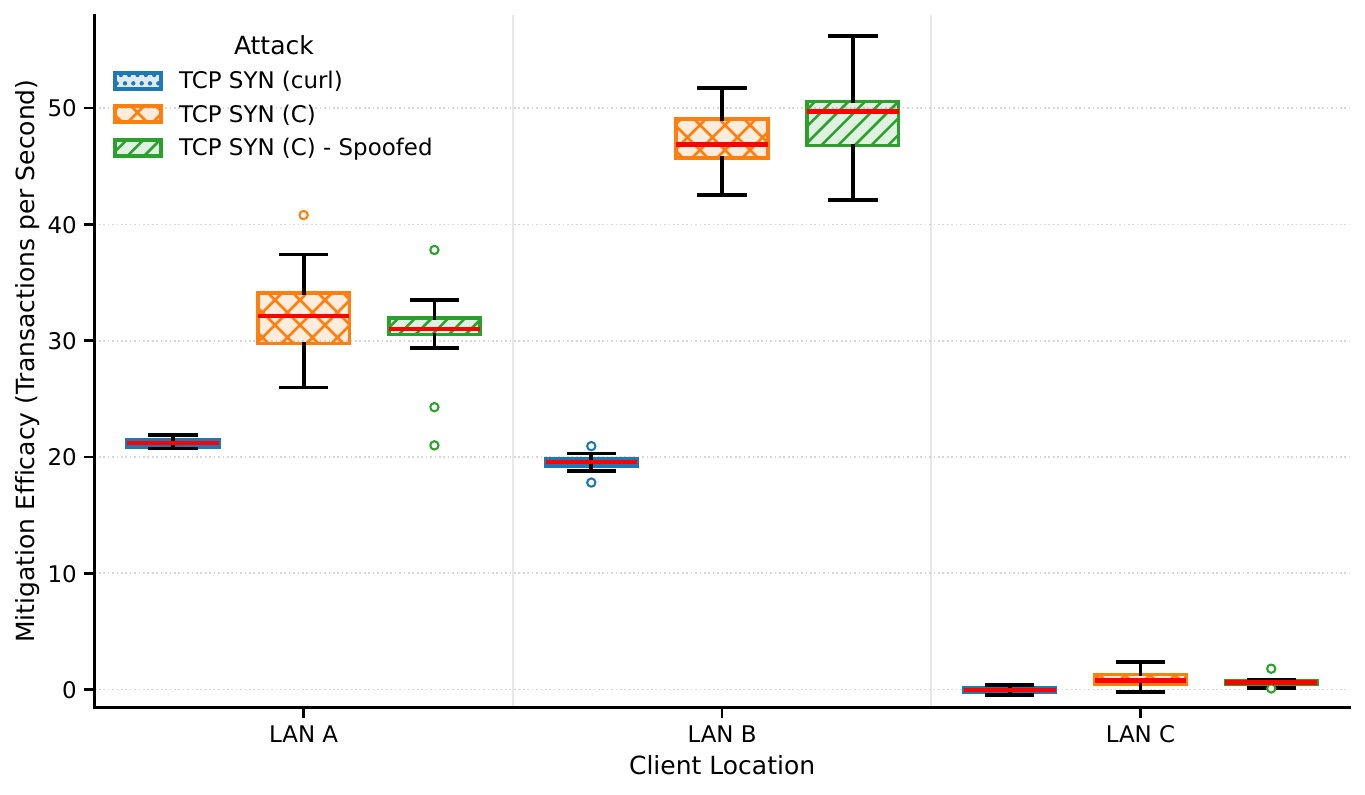}
    \caption{EdgePoW restores QoS via ingress-aware edge filtering.}
  \end{subfigure}
  \caption{Experimental results comparing defense mechanisms.
    (a)~Both SYN Cookies and EdgePoW impose negligible
    peacetime overhead. (b)~Without defense, the 350\,Kpps
    flood causes complete DoS for LAN~C and severe degradation
    elsewhere. (c)~SYN Cookies worsen congestion by generating
    per-SYN responses. (d)~EdgePoW raises difficulty on the
    attacked ingress and restores QoS network-wide.}
  \label{fig:results}
\end{figure*}

Baseline QoS (no attack, no defense) is $\approx$55\,TPS for
LAN~A (co-located with the server) and $\approx$51\,TPS for
LANs~B and~C (traversing one and two hops respectively).
Peacetime overhead with EdgePoW at $d_\mathit{default}=4$
is ${<}\,2\%$, comparable to SYN Cookies (Fig.~\ref{fig:results}a).
Under a 350\,Kpps spoofed flood with no defense, LAN~C suffers
complete denial of service (0\,TPS) and LANs~A/B experience
severe degradation, losing 40--60\% of their baseline
throughput (Fig.~\ref{fig:results}b). This confirms that
volumetric SYN floods cause network-wide collateral damage
beyond the target ingress.

\subsection{Defensive Efficacy}

\paragraph{SYN Cookies.}
Under the spoofed flood, SYN Cookies yield negative efficacy
(Fig.~\ref{fig:results}c): every SYN elicits a SYN-ACK,
doubling egress load and worsening congestion for all LANs.
LAN~C throughput drops further, and even LAN~A degrades more
than without any defense.

\paragraph{EdgePoW}
Our framework achieves strong positive efficacy
(Fig.~\ref{fig:results}d). The controller pushes a
high-difficulty rule to $r_2$; under spoofed floods the
ingress-wide rule persists, under stable-source floods it
narrows to the attacker /24. Invalid SYNs are dropped before
consuming core bandwidth, so LANs~A/B rebound to near-baseline
QoS while LAN~C benign clients maintain service.

\subsection{Ingress-First, Prefix-Refined Enforcement}

\begin{table}[t]
  \caption{Effect of enforcement granularity at the contested
    ingress $r_2$. Attack suppression remained above 96\% in
    all three settings.}
  \label{tab:granularity}
  \centering
  \footnotesize
  \resizebox{\columnwidth}{!}{%
  \begin{tabular}{p{2.5cm}p{2.4cm}cc}
    \toprule
    Scenario & Final rule & \makecell{Benign\\LAN C TPS} & \makecell{Median\\setup delay} \\
    \midrule
    Stable-source flood, ingress-only
      & $(r_2, \mathrm{SYN})$ & 41.8 & 2.6\,s \\
    Stable-source flood, refined /24
      & $(r_2, \text{attacker /24}, \mathrm{SYN})$ & 46.7 & 1.7\,s \\
    Spoofed randomized-source flood
      & $(r_2, \mathrm{SYN})$ & 42.4 & 2.3\,s \\
    \bottomrule
  \end{tabular}}
\end{table}

Table~\ref{tab:granularity} compares enforcement granularity
on ingress $r_2$, where attacker and benign clients share the
same ingress but occupy different /24 prefixes. Ingress-only
enforcement preserves most QoS for benign LAN~C clients;
however, prefix refinement further recovers an additional
4.9\,TPS ($+11.7\%$) and reduces median connection setup
latency from 2.6\,s to 1.7\,s by exempting the benign /24
from elevated difficulty. Under spoofed floods with randomized
source IPs, no single prefix dominates, so the controller
correctly retains the coarser ingress-wide rule. Attack
suppression remained above 96\% in all three configurations.

\subsection{Impact on Legitimate Users}

Legitimate-client cost remains modest:
$T_\mathit{conn}=2^d/H$. At $d=16$, a desktop-class host
($H\approx2{\times}10^8$ hashes/s) spends $\approx0.3$\,ms,
while an IoT-class device ($H\approx10^5$ h/s) spends
$\approx0.65$\,s. DDP needs at most four retries after a policy
change; later connections reuse the confirmed cache.

\subsection{DDP Robustness}
\label{sec:ddp-robustness}

Table~\ref{tab:ddp} shows DDP is conservative: under 1--2\%
random loss, fewer than 0.8\% of connections transiently probe
higher difficulty and none cache it. Attack onset converges from
$d=4$ to $d=16$ in three retries (3.1\,s median), attack
clearance succeeds immediately, and server-unreachable failures
leave the cache unchanged.

\begin{table}[t]
  \caption{DDP robustness across 10,000 connection attempts per
    scenario. Transient escalations are not cached.}
  \label{tab:ddp}
  \centering
  \begingroup
  \renewcommand{\arraystretch}{0.92}
  \resizebox{\columnwidth}{!}{%
  \begin{tabular}{lcccc}
    \toprule
    Scenario & Success & \makecell{Median\\retries} & \makecell{Median\\setup} & Outcome \\
    \midrule
    Steady $d=4$, no loss                       & 99.9\% & 0 & 23\,ms & stable cache \\
    Steady $d=4$, 1\% loss                      & 99.7\% & 0 & 39\,ms & 0.2\% transient, 0 cached \\
    Steady $d=4$, 2\% loss                      & 99.3\% & 0 & 67\,ms & 0.7\% transient, 0 cached \\
    Steady $d=4$, $4{\times}$ flash crowd       & 98.9\% & 0 & 81\,ms & 0.4\% transient, 0 cached \\
    Attack onset: cache $4 \to \text{network}\ 16$      & 98.6\% & 3 & 3.1\,s & converges to $d=16$ \\
    Attack clear: cache $16 \to \text{network}\ 4$      & 99.9\% & 0 & 24\,ms & stale high $d$ accepted \\
    Server unreachable                            & 0.0\%  & 4 & 4.0\,s & cache unchanged \\
    \bottomrule
	  \end{tabular}}
  \endgroup
\end{table}

\subsection{Detection Parameter Sensitivity}
\label{sec:sensitivity}

Table~\ref{tab:sensitivity} summarizes the
responsiveness/FPR trade-off. One-second detection catches
floods in $\approx1.1$--$1.2$\,s but can raise FPR to 4.8\%;
a 10\,s window removes false positives but lets 3.5\,M SYNs
through. We use $\theta=3,\tau_\mathit{detect}=3$\,s: zero FPR,
$\approx3.1$\,s detection, and 1.1\,M exposed SYNs.

\begin{table}[t]
  \caption{Detection parameter sensitivity ($\beta=0.8$,
    $\tau_\mathit{clear}=10$\,s). Attack rate
    $\approx 350$\,Kpps from LAN~C.}
  \label{tab:sensitivity}
  \centering
  \footnotesize
  \begingroup
  \renewcommand{\arraystretch}{0.92}
  \resizebox{\columnwidth}{!}{%
  \begin{tabular}{ccccc}
    \toprule
    $\theta$ & $\tau_\mathit{detect}$ (s) & $\bar{T}_\mathit{det}$ (s) &
      FPR (\%) & \makecell{Unmitigated SYNs\\(${\times}10^3$)} \\
    \midrule
    1.5 & 1  & $1.2 \pm 0.3$ & 4.8 & 420   \\
    2.0 & 1  & $1.1 \pm 0.2$ & 1.6 & 385   \\
    3.0 & 1  & $1.1 \pm 0.2$ & 0.2 & 385   \\
    5.0 & 1  & $1.0 \pm 0.1$ & 0.0 & 350   \\
    2.0 & 3  & $3.2 \pm 0.4$ & 0.3 & 1,120 \\
    \textbf{3.0} & \textbf{3} & $\mathbf{3.1 \pm 0.3}$ & \textbf{0.0} & \textbf{1,085} \\
    3.0 & 5  & $5.1 \pm 0.3$ & 0.0 & 1,785 \\
    3.0 & 10 & $10.1 \pm 0.2$& 0.0 & 3,535 \\
    \bottomrule
	  \end{tabular}}
  \endgroup
\end{table}

\subsection{Switch Verification Overhead}

Table~\ref{tab:switch} reports switch-side verification cost.
A single core verifies 20.8\,Mpps versus the 350\,Kpps attack
rate, and OVS throughput drops only 2.8\% with PoW enabled.
Verification performs one hash per SYN, so cost is independent
of difficulty $d$.

\begin{table}[t]
  \caption{Switch-side PoW verification microbenchmarks.}
  \label{tab:switch}
  \centering
  \small
  \begin{tabular}{lc}
    \toprule
    Metric & Value \\
    \midrule
    Single verification latency & $48 \pm 3$\,ns \\
    Max verification throughput (1 core) & 20.8\,Mpps \\
    OVS base forwarding rate (no PoW)    & 2.14\,Mpps \\
    OVS forwarding rate (PoW on SYNs)    & 2.08\,Mpps \\
    Throughput reduction                 & 2.8\% \\
    \bottomrule
  \end{tabular}
\end{table}

%%
%% 6. DISCUSSION
%%
\section{Discussion}

EdgePoW treats unstable sources conservatively: prefix spreading
or spoofing keeps ingress-wide enforcement in place, while
low-rate traffic stays at peacetime difficulty. Timestamp
binding ($\delta=30$\,s) limits precomputation to $\approx$11\,M
proofs per hour, and $d=16$ caps each node at $\approx$3\,K
compliant SYNs/s. The design assumes managed programmable
edges; client PoW can run as an eBPF socket hook~\cite{delaughter2025syn},
non-participants incur no cost when $d_\mathit{default}=0$, and
prefix refinement stays local to each ingress. EdgePoW mitigates
volumetric SYN floods rather than application-layer DoS, and
non-participating clients need PoW support only during active floods.

%%
%% 7. RELATED WORK
%%
\section{Related Work}

SYN Cookies~\cite{pan2024tcp,delaughter2025syn} prevent state exhaustion but
can amplify bandwidth by responding to every SYN~\cite{dean2001client,yoo2024smartcookie};
SmartCookie~\cite{yoo2024smartcookie} moves this logic into P4
switches but still sends per-SYN responses. Interactive PoW
defenses~\cite{juels1999client,noureddine2019revisiting} consume server egress
for challenges, and SDN-based DDoS defenses~\cite{shao2023,jia2026blocksdnvc}
often rely on blacklists that risk collateral damage under
shared or spoofed addresses. EdgePoW instead combines
non-interactive SYN PoW, ingress-aware enforcement, and prefix
refinement; industry reports~\cite{cloudflare2024,yoachimik2024q1,yoachimik2024q4}
confirm that SYN floods remain a dominant DDoS vector.

%%
%% 8. CONCLUSION
%%
\section{Conclusion}

EdgePoW combines non-interactive PoW, ingress-aware SDN
control, and success-confirmed difficulty discovery for
volumetric TCP SYN floods. Experiments show restored
application QoS where SYN Cookies worsen congestion, 11.7\%
higher benign-client throughput than ingress-only enforcement,
and below 0.8\% transient false escalations under 2\% packet
loss. Future work includes P4 verification and larger-scale evaluation.

\begin{acks}
This work was supported by the New Generation Artificial Intelligence
National Science and Technology Major Project (No.~2025ZD0122304)
and the Shenzhen Science and Technology Program
(Nos.~ZDCY20250901095900001 and SYSPG20241211173440003). We used
Grammarly for grammar corrections and ChatGPT for language polishing.
\end{acks}

%%
%% REFERENCES
%%
\bibliographystyle{ACM-Reference-Format}

\end{document}